\documentclass[galaxies,article,accept,moreauthors,pdftex,10pt,a4paper]{mdpi} 

\firstpage{1} 
\pubvolume{xx}
\issuenum{1}
\articlenumber{1}
\pubyear{2018}
\copyrightyear{2018}
\externaleditor{Academic Editor: name}
\history{Received: date; Accepted: date; Published: date}

\newcommand{\apj}{Astrophysical Journal}
\newcommand{\apjl}{Astrophysical Journal Letters}
\newcommand{\prd}{Physical Review D}
\newcommand{\aap}{Astronomy and Astrophysics}
\newcommand{\mnras}{Monthly Notices of the Royal Astronomical Society}
\newcommand{\nat}{Nature}
\newcommand{\ssr}{Space Science Reviews}

\pdfoutput=1

\Title{IMAGINE: modeling the Galactic magnetic field}

\Author{Marijke Haverkorn $^{1}$\orcidA{}, 
Fran\c{c}ois Boulanger $^2$, 
Torsten En\ss lin $^3$, J\"org R. H\"orandel $^{1,4}$, Tess Jaffe $^{5,6}$, 
Jens Jasche $^7$, J\"org P. Rachen $^1$ and Anvar Shukurov $^8$}

\AuthorNames{Marijke Haverkorn et al}

\address{%
$^{1}$ \quad Department of Astrophysics/IMAPP, Radboud University,
P.O. Box 9010, 6500 GL Nijmegen,
The Netherlands; m.haverkorn@astro.ru.nl\\
$^{2}$ \quad LERMA/LRA, Observatoire de Paris, PSL Research University, CNRS, Sorbonne Universit\'e,
UPMC Universit\'e Paris 06, Ecole Normale Sup\'erieure, 75005 Paris,
France\\
$^{3}$\quad Max Planck Institute for Astrophysics, Karl-Schwarschildstr. 1, 85741 Garching, Germany\\
$^{4}$\quad Nikhef, Science Park, Amsterdam, The Netherlands\\
$^{5}$\quad CRESST II, NASA Goddard Space Flight Center, Greenbelt, MD, 20771, USA\\
$^{6}$\quad Department of Astronomy, University of Maryland, College Park, MD, 20742, USA\\
$^{7}$\quad The Oskar Klein Centre, Department of Physics, Stockholm University, AlbaNova University Centre,
SE 106 91 Stockholm, Sweden\\
$^{8}$\quad School of Mathematics, Statistics and Physics, Newcastle University, Newcastle upon Tyne,
NE1 7RU, UK}

\abstract{The IMAGINE Consortium aims to bring modeling of the magnetic
  field of the Milky Way to a next level, by using Bayesian
  inference. IMAGINE includes an open-source modular software pipeline
  that optimizes parameters in a user-defined Galactic magnetic field
  model against various selected observational datasets. Bayesian
  priors can be added as external probabilistic constraints of the
  model parameters. These conference proceedings describe the science
  goals of the IMAGINE Consortium, the software pipeline and its
  inputs, viz observational data sets, Galactic magnetic field models,
  and Bayesian priors.}


\begin{document}

\section{Introduction}

The magnetic field in the Milky Way is an invisible agent controlling
many physical processes, such as carving gas clouds into filaments,
delaying star formation, accelerating and guiding cosmic rays, shaping
supernova remnants, and puffing up the Galactic halo. The energy
density contained in these magnetic fields is comparable to that in
stars, in cosmic rays and in the turbulent gas, causing constant
interaction with and feedback from all these components. Therefore,
understanding the Galactic ecosystem is a major science goal for
studies of the magnetic field in the Milky Way. In addition, the shape
and strength of magnetic fields in galaxies provides clues as to the
origin and the evolution of galactic magnetic field, which in turn can
constrain our ideas of formation of galaxies and large-scale structure
in the early Universe. Lastly, the Galactic magnetic field can also be
a hindrance to studies wanting to investigate what lies beyond: it
muddles our view of sources of Ultra-High Energy Cosmic Rays and of
extragalactic magnetic fields, it causes small-scale polarized
synchrotron radiation that can mimic the signature of the 21cm line of
neutral hydrogen in the early Universe's Dark Ages, and produces dust
polarization that hampers Cosmic Microwave Background polarization
measurements.  In summary, the magnetic field of the Milky Way is
important for a broad range of research areas, whether it will be used
to study it or subtract it.

Observational and theoretical knowledge of magnetic fields in the
Milky Way still gives us a sketchy image. The idea that a magnetic
dynamo has amplified and maintains the Galactic magnetic field is
well-established. However, it is far from clear what the most
important driving agents of this dynamo are, e.g.\ turbulence
or cosmic-ray buoyancy. Large-scale mean-field dynamo models can
be described analytically as a combination of modes, some of which are
more likely to be excited than others in disk galaxies. The
small-scale magnetic field structure is often modeled as Gaussian
turbulence, which is too simplified since magnetic turbulence is highly
anisotropic and intermittent.  There are a large number of indirect
observational tracers for magnetic fields in the Galaxy: these probe
either the strength or direction of the field, either the component
parallel to the line of sight or the component in the plane of the
sky. In addition, these tracers depend on the conditions in the probed
region, making reconstruction of the total field strength and
direction at a certain location all but impossible.

Much research has been done in the past decades trying to construct a
Galactic magnetic field model that is consistent with the
observational tracers. However, these studies are very heterogeneous
in the sense that they use different parameterizations of the Galactic
magnetic field, and a different subset of observational tracers. This
makes it impossible to quantitatively assess their quality and to
intercompare them. The Interstellar MAGnetic field INference Engine
(IMAGINE) research project aims to take the next step in Galactic
magnetic field modeling by including all possible observational
tracers and use Bayesian inference techniques, allowing the inclusion
of prior (observational and theoretical) knowledge on galactic
magnetism, and quantifiable quality assessment of different models.

This proceedings contribution briefly discusses the aims and methods of the
IMAGINE Consortium. A much more elaborate discussion about the content
and science goals of IMAGINE can be found in the IMAGINE White Paper
\cite{boulangeretal18}.

%%%%%%%%%%%%%%%%%%%%%%%%%%%%%%%%%%%%%%%%%%%%%%%
\section{Introducing IMAGINE}

\subsection{What is IMAGINE?}
IMAGINE is both a consortium and a software package. The consortium is
an international, broad, and open group of researchers who are joined
by an interest in galactic magnetic fields.  The goal of the
consortium is to coordinate and facilitate the development and
improvement of models for the magnetic field in the Milky Way, with
the broader goal of gaining more insight in a wide variety of science
questions that are influenced by galactic magnetic fields in various
ways. 

The means of achieving this is the development and use of the IMAGINE
software, which models the Galactic magnetic field using Bayesian
inference and all possible observational inputs and priori knowledge
on the Galactic magnetic field. The basic IMAGINE pipeline is written
and tested \cite{steiningeretal18}, see Section~\ref{s:pipeline}.

\begin{figure}[t]
\centering
\includegraphics[width=\textwidth]{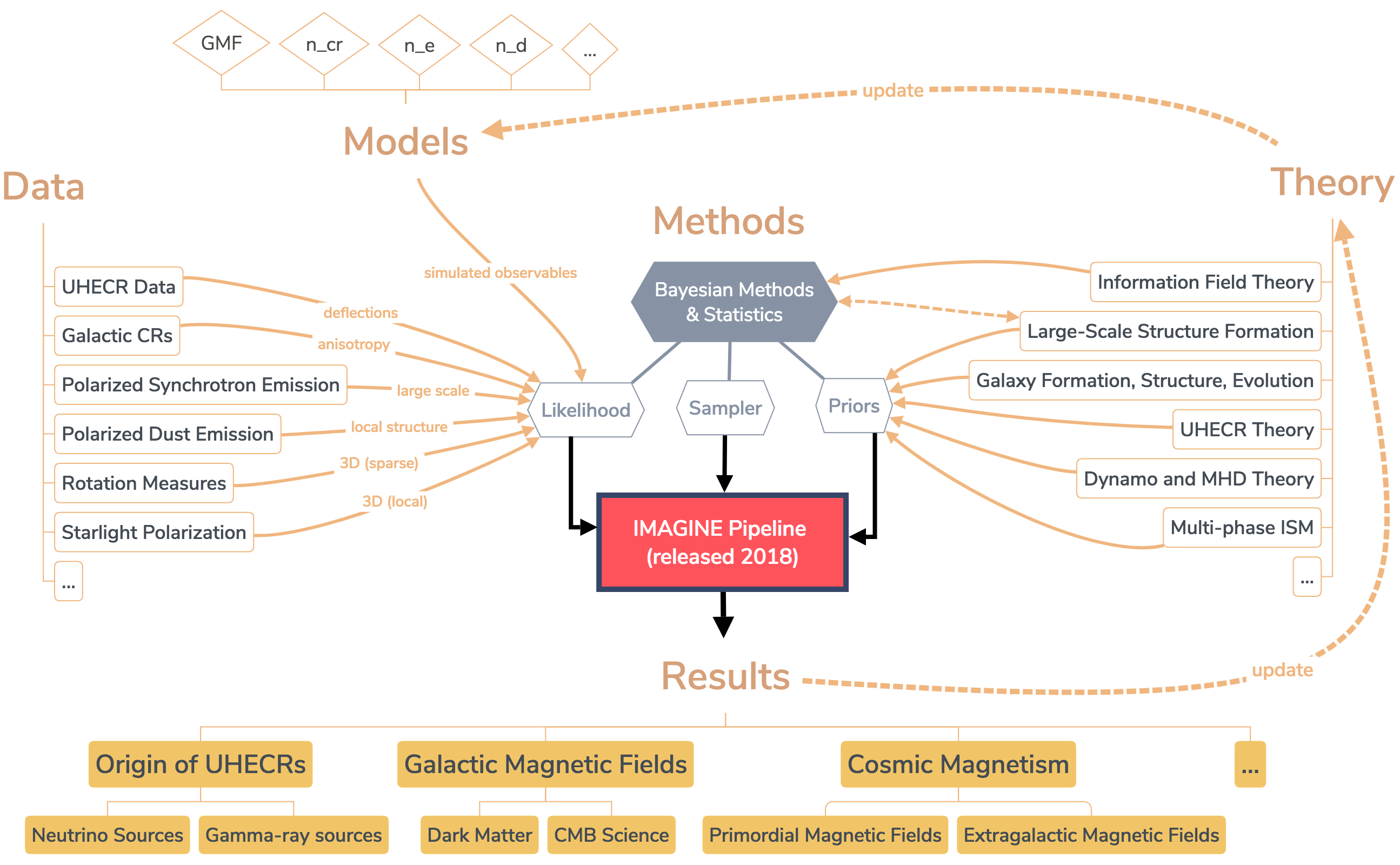}
\caption{A mind map of the IMAGINE project \cite{boulangeretal18}.}
\end{figure}

\subsection{Scientific aims of IMAGINE}

Knowledge of the strength and structure of the Galactic magnetic field
is indispensable for studies of the Milky Way's interstellar medium
and cosmic rays. It is valuable input for dynamo theories and studies
of galactic and extragalactic magnetism. In addition, the magnetic
field of the Milky Way is a foreground influencing various
extragalactic studies such as Ultra-High Energy Cosmic Ray (UHECR)
sources, Cosmic Microwave Background polarization or the Epoch or
Reionization.  

The main goal of the IMAGINE Consortium is modeling of the Galactic
magnetic field. Major science questions that can be addressed with
sufficiently reliable knowledge of the Galactic magnetic field are the
following:

\paragraph{\it{What is the role of the Galactic magnetic field in the
  interstellar medium?}}
It is widely known that Galactic magnetic fields are dynamically
important, and that their energy density is comparable to other
components in the Milky Way \cite{heileshaverkorn12}. Recently, the
close coupling between these components has become clear in detail:
hydrogen filaments are aligned in magnetic fields in dust as probed by
starlight polarization \cite{mccluregriffithsetal05}, by polarized
dust emission \cite{clarketal14} and by Faraday rotation \cite{kalberlakerp16,
  kalberlaetal17}. State-of-the-art observational data sets such as
Planck polarization maps \cite{planck15VII}, all-sky Faraday rotation measure
maps \cite{oppermannetal12}, and many HI and (polarized) synchrotron
studies with current and next-generation instruments necessitate
detailed knowledge on the turbulent components of the Galactic
magnetic field to ensure progress.

\paragraph{\it{How are Galactic cosmic rays accelerated and propagated in
  the Galactic magnetic field?}}
Electrons, positrons and ions are accelerated under the influence of
small-scale magnetic fields to cosmic rays. These cosmic rays
propagate through the Galaxy mostly along field lines, but also by
diffusion and advection. They interact with the interstellar
environment through e.g.\ spallation or gas heating; understanding
these processes, and the creation and propagation of Galactic cosmic
rays therefore needs accurate knowledge about the local magnetic field
strength and structure.

\paragraph{\it{How are magnetic fields in galaxies amplified and
  maintained?}}
There are a number of physical processes that can create very weak magnetic
fields in the early Universe. These fields are then amplified by a
magnetic dynamo mechanism to the microGauss strengths we observe in
current galaxies. How that magnetic dynamo works in detail and what drives it
are as yet open questions. Various dynamo models give different
predictions for (symmetries) in the large-scale structure of Galactic
magnetic fields. Hence, a good model of the Galactic magnetic field
can constrain these theories and shed light on the origin and
evolution of galactic magnetism.

\paragraph{\it{What are the sources of ultra-high energy cosmic rays?}}
Ultra-high energy cosmic rays (UHECRs) are extragalactic cosmic rays
with energies exceeding $E \sim 10^{18}$~eV. Various cosmic sources
could possibly accelerate charged particles to these extreme energies,
such as compact transients connected to star forming activity in
galaxies, AGN, radio galaxies or galaxy clusters, but the relative
contribution of these sources is still unknown. UHECRs propagating
from their sources through the magnetized intergalactic and
interstellar medium to detectors at Earth are deflected by the
intergalactic and (in many directions dominant) Galactic magnetic
fields. The deflections in these magnetic fields currently preclude us
from following these particles back to their sources. A sufficiently
reliable model of the (large-scale) magnetic field of the Galaxy would
enable us to determine the sources of the UHECRs.

Apart from these science questions, knowledge of the magnetic field
and non-thermal components of the Milky Way can be useful more
tangentially in the study of galaxy formation. One of the most
stringent problems in galaxy formation studies is the ``missing
satellite problem'', i.e.\ the problem that numerical simulations
predict many more low-mass satellite galaxies than are observed. This
discrepancy can possibly be attributed to the underestimation of feed
back processes by stellar winds, supernova remnants and active
galactic nuclei in these simulations. This feedback could drive gas
out of galaxies, or preclude new gas from accreting onto the
galaxies. Studies of the non-thermal components of the Milky Way will
teach us about these feedback processes at low star formation rates.

Finally, a trustworthy model of the Galactic magnetic field will allow us
to properly subtract the (polarized) radiation fields in the Galaxy
that arise due to this magnetic field, to reveal the extragalactic sky
in more detail. We will be able to better model Galactic polarized
dust emission to benefit detection of polarized Cosmic Microwave
Background B-modes, to model polarized Galactic synchrotron emission
to aid detection of HI fluctuations from the Epoch of Reionization, or
to study magnetic fields in the cosmic web.

%%%%%%%%%%%%%%%%%%%%%%%%%%%%%%%%%%%%%%%%%%
\section{The IMAGINE software package}
\label{s:pipeline}

\begin{figure}[t]
\centering
\includegraphics[width=0.7\textwidth]{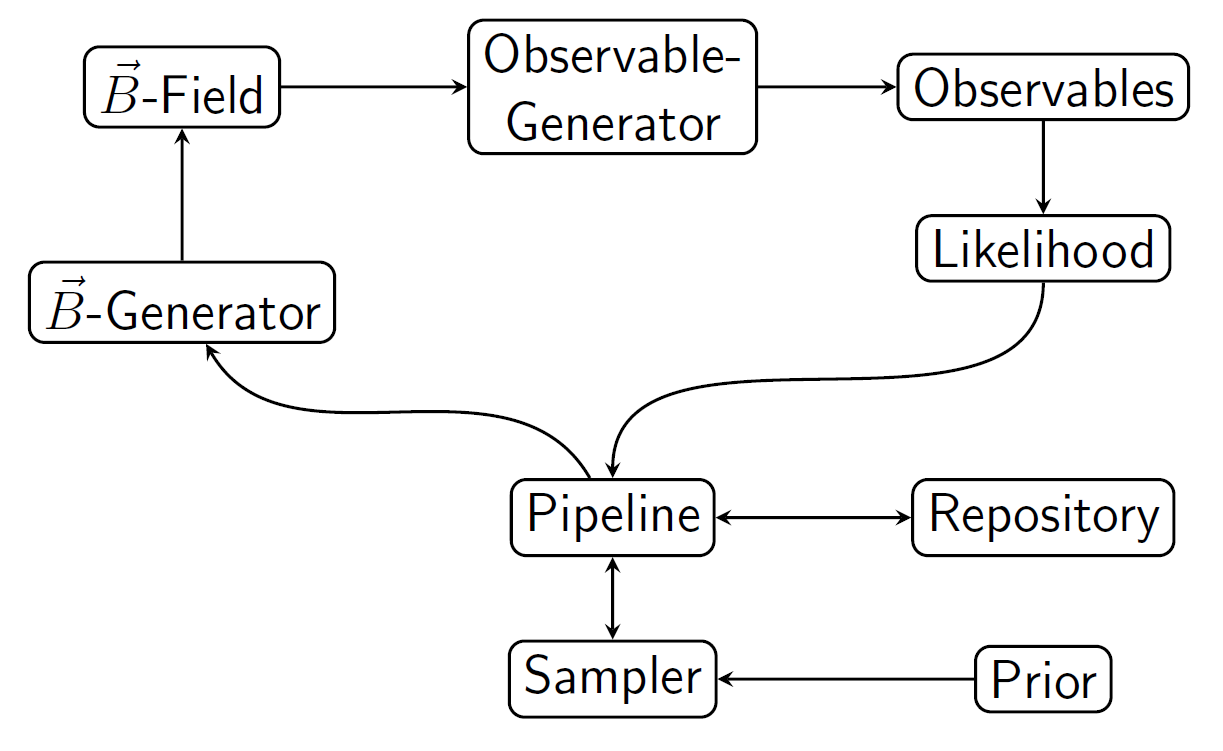}
\caption{The structure of the IMAGINE project: generating models,
  comparison to data and sampling parameter space \cite{boulangeretal18}.}
\label{f:structure}
\end{figure}

The IMAGINE software is a modular open source framework for doing inference
on generic parametric models of the Galaxy
\cite{steiningeretal18}. Figure~\ref{f:structure} shows its general
outline.

Within the IMAGINE pipeline, a magnetic field configuration is
generated according to a user-defined model (see
Section~\ref{s:gmfmodels}). Using this magnetic field configuration,
combined with other necessary input models such as thermal electron
density, cosmic ray electron density, and dust
distributions, all-sky maps of observables are generated using the
Hammurabi software package \cite{waelkensetal09}. The generated
observables can be synchrotron intensity, synchrotron polarized
intensity, polarized dust emission or rotation measure, depending on
the available data. Comparison of the generated data with
observational maps (Section~\ref{s:obs}) results in likelihood
evaluations for the magnetic field parameters, which are saved in a
repository, after which a sampler selects a new set of parameters in
the magnetic field model, using information from priors
(Section~\ref{s:priors}) if available.

This framework ensures significant improvement of this project over
earlier Galactic magnetic field modeling in literature in multiple ways:
\begin{itemize}
\item Priors: any knowledge that we already have about
  Galactic magnetic fields can be included in the modeling;
\item Addition of observational tracers: the modular set-up of IMAGINE
  enables inclusion of different and heterogeneous data sets in a
  consistent way;
\item Evidence: the Bayesian evidence component in IMAGINE allows
  quantitative comparison of various Galactic magnetic field models in
  the literature. Apart from comparison of various models using the
  same data by the same authors, comparison of models in different
  publications using different data sets is only attempted by the
  Planck collaboration \cite{planckxlii}; quantitative comparison is
  currently completely lacking.
\end{itemize}

%%%%%%%%%%%%%%%%%%%%%%%%%%%%%%%%%%%%%%%%%%
\section{Observational tracers}
\label{s:obs}

The ambitious goal of IMAGINE is to, eventually, use all possible
observational tracers. This includes first of all the 'traditional'
observational tracers, i.e.\ tracers that are currently used for model
fitting of the Galactic magnetic field (see e.g. \cite{beck15, han17}). These are:
\begin{itemize}
\item (Polarized) synchrotron emission. Synchrotron emission traces
  the magnetic field component perpendicular to the line of sight,
  integrated over the line of sight. Its polarization can be used as a
  measure of the ratio of the turbulent to the uniform magnetic field
  components.
\item Faraday Rotation of polarized radio emission from
  pulsars. Pulsar Rotation Measures (RMs) give information on the magnetic
  field component parallel to the line of sight $B_{\parallel}$,
  weighted by the electron density $n_e$: RM~$\propto \int n_e
  B_{\parallel} ds$. Although these data have the great advantage that
  they probe magnetic field in the 3D volume of the Galaxy (instead of
  most other tracers that integrate over pathlengths through the
  entire Galaxy), pulsar RMs are fairly scarce and mostly concentrated
  near the Galactic plane. 
\item Rotation Measures from extragalactic point sources. Mostly
  thanks to the RMs from the all-Northern-sky NVSS survey
  \cite{tayloretal09}, the RM Grid from extragalactic sources now
  contains $\sim 40,000$~sources  across the whole sky, from which a
  Faraday Rotation map of the sky can be constructed \cite{oppermannetal12}.
\item Thermal emission from dust is partially polarized due to the
  influence of magnetic fields. All-sky dust polarization maps
  from e.g.\ the Planck satellite, combined with a dust model, give an
  independent measure of the averaged magnetic field component perpendicular to
  the line of sight.
\end{itemize}

In addition to these tracers, there are a number of observational
tracers that have so far been used little, if at all, to probe the
large-scale magnetic field due to great complexity or
uncertainties. However, improved technological capabilities are making
it worthwhile to start thinking about these tracers. They are:
\begin{itemize}
\item Starlight polarization: optical and near-infrared starlight gets
  partially polarized by elongated dust grains rotating aligned with
  an interstellar magnetic field. Small fields of view have been used
  to constrain Galactic magnetic field models
  \cite{paveletal12}. Future all-sky optical polarimetric surveys
  \cite{magalhaes14} combined with GAIA distances and reliable dust
  models will be included into IMAGINE. The short lines of sight to
  the stars will make this data set very complementary to the
  traditional tracers that integrate along the line of sight through
  the entire Galaxy.
\item Galactic cosmic ray distribution: small-scale anisotropies in
  Galactic TeV cosmic ray arrival directions have recently been
  discovered by several experiments \cite{amenomorietal10,
    aartsenetal16}. In principle, this can be predicted in the Galaxy
  model generated in IMAGINE, which can provide additional constraints
  to the model parameters.
\item UHECR arrival directions: UHECRs are deflected in the Galactic
  magnetic field, for the highest measured energies typically by a few
  degrees to a few tens of degrees, depending on their rigidity and
  propagation path. A Bayesian optimization of the source
  distribution(s) of UHECRs as well as their deflections in intergalactic
  space and in the Galaxy against their arrival directions at Earth
  can be used to constrain possible Galactic magnetic field
  configurations.

\item Faraday Tomography: in principle, Faraday Tomography maps of the
  IMAGINE Galaxy model could be compared to observed Faraday
  Tomography maps. Faraday Tomography at low (LOFAR, MWA) frequencies
  reveals more about discrete structures in the local neighborhood and
  therefore may not be ideal. However, at higher (GHz) frequencies,
  Faraday Depth spectra likely better represent the global Galactic
  magnetic field and as such can be used to constrain model
  parameters.
\end{itemize}

%%%%%%%%%%%%%%%%%%%%%%%%%%%%%%%%%%%%%%%%%%
\section{Galactic magnetic field models}
\label{s:gmfmodels}

A large number of parameterized, heuristic models of the
Galactic magnetic field exist, which are heterogeneous in many
different ways. These models fit the Galactic disk only (e.g.\
\cite{vanecketal11}) or fit both disk and halo magnetic fields. They
include different magnetic field components, such as regular and
turbulent magnetic fields, anisotropic turbulent magnetic fields
(e.g. \cite{jaffeetal10}), or vertical magnetic fields
(e.g. \cite{jf12a, jf12b}). They can use either simple Gaussian slabs
to estimate the underlying cosmic-ray density, or complex models like
GALPROP \cite{strongmoskalenko98} based on observational data
\cite{jaffeetal11}. Also, they are based on various observational data
sets, such as (a subset of) extragalactic source rotation measures,
pulsar rotation measures (e.g.\ \cite{menetal08}), synchrotron
intensity, synchrotron polarization (e.g.\ \cite{sunetal08}),
thermal dust emission (e.g. \cite{fauvetetal11}), and thermal dust
polarization (e.g. \cite{jaffeetal13}) (see
\cite{haverkorn15} for a review).

Using IMAGINE, it becomes relatively straightforward to quantify how
well each of these models fit to the data by computing the Bayesian
evidence. This will take into account possible varying observational
data and varying numbers of free parameters. Also, all models can be
compared fitting against the same observational data.

As a next step, we intend to include Galactic magnetic field
models based on physical coherent field configurations
(e.g. \cite{ferriereterral14}) or including non-Gaussian turbulent
fields \cite{shukurovetal17}. Eventually, the goals include exploring
non-parametric modeling of the Galactic magnetic field, in which the
parameters are treated as fields optimized under certain prior
constraints (see e.g. \cite{ensslinetal09}) .

%%%%%%%%%%%%%%%%%%%%%%%%%%%%%%%%%%%%%%%%%%
\section{Bayesian priors}
\label{s:priors}

Any existing knowledge of the Galactic magnetic field can be coded as
a Bayesian prior, and used as a constraint in the model fitting. 

Prior information not used in the heuristic Galactic magnetic field
models described above are the equations describing the physical
process underlying the magnetic field configuration, i.e.\ the dynamo
equations. Any magnetic field can be described as a superposition of
various dynamo modes, which are eigenfunctions of the mean-field
dynamo equation. As it is much more likely to observe a growing dynamo
mode than a decaying one, the probabilities of these eigenvalues of
the dynamo modes can be used as prior information in the modeling.

In addition, any constraints on the Galactic magnetic field known from
previous (unrelated) observational data can be used as a prior, such
as information on the magnetic field in the local Solar neighborhood
(e.g.\ \cite{alvesetal18}) or about large-scale field reversals in the
extended Solar neighborhood (e.g.\ \cite{simardnormandinkronberg79}).
Magnetic fields are measured in dozens of nearby spiral galaxies
\cite{beck15}, which together provide rich constraints on
probabilities of various galactic magnetic field configurations.

Possible more general priors are based on cosmic evolution of galactic
magnetic fields, by combining galactic magnetic field measurements in
the early Universe \cite{farnesetal14} with cosmological simulations of
galaxy formation.

Finally, priors can be constructed following the original idea behind
the IMAGINE Consortium: combining knowledge on UHECRs and Galactic
magnetism. The sources of UHECRs are still not known, but the many
observations have narrowed down the possibilities to a few physically
plausible source categories. These can be assigned probabilities,
after which observed UHECRs arrival directions at Earth can be used to
constrain Galactic magnetic field models.

%%%%%%%%%%%%%%%%%%%%%%%%%%%%%%%%%%%%%%%%%%
\section{Conclusions}

IMAGINE is a global effort to take modeling of Galactic magnetic
fields to the next level.  There is a multitude of science questions
in many different fields of astrophysics that could benefit from the
usage of the IMAGINE software. The current IMAGINE Consortium is
addressing some of these, but there is still a lot of uncharted
terrain. The consortium encourages usage of the software pipeline and
welcomes new members.

\acknowledgments{We thank the International Space Science Institute in
  Bern, Switzerland, and the Lorentz Center in Leiden, the
  Netherlands, for the hospitality and financial support that has led
  to the founding of the IMAGINE Consortium.}


%=====================================
% References, variant A: internal bibliography
%=====================================
\reftitle{References}

% The following MDPI journals use author-date citation: Arts, Econometrics, Economies, Genealogy, Humanities, IJFS, JRFM, Laws, Religions, Risks, Social Sciences. For those journals, please follow the formatting guidelines on http://www.mdpi.com/authors/references
% To cite two works by the same author: \citeauthor{ref-journal-1a} (\citeyear{ref-journal-1a}, \citeyear{ref-journal-1b}). This produces: Whittaker (1967, 1975)
% To cite two works by the same author with specific pages: \citeauthor{ref-journal-3a} (\citeyear{ref-journal-3a}, p. 328; \citeyear{ref-journal-3b}, p.475). This produces: Wong (1999, p. 328; 2000, p. 475)

%=====================================
% References, variant B: external bibliography
%=====================================
%\externalbibliography{yes}
%\bibliography{your_external_BibTeX_file}

\end{document}